\documentclass[pra,twocolumn,showpacs,preprintnumbers,amsmath,amssymb,superscriptaddress]{revtex4}

\usepackage{graphicx}
\usepackage{natbib}
\usepackage{latexsym}
\begin{document}

\newcommand{\ha}{\hat{a}}
\newcommand{\ddt}{\frac{d}{dt}}
\newcommand{\ddtau}{\frac{d}{d\tau}}
\newcommand{\hA}{\hat{A}}
\newcommand{\hB}{\hat{B}}
\newcommand{\hC}{\hat{C}}
\newcommand{\hD}{\hat{D}}
\newcommand{\hS}{\hat{S}}
\newcommand{\tA}{\tilde{A}}
\newcommand{\tB}{\tilde{B}}
\newcommand{\tC}{\tilde{C}}
\newcommand{\tD}{\tilde{D}}
\newcommand{\dnor}{\Delta^n}
\newcommand{\dan}{\Delta^a}
\newcommand{\hn}{\hat{n}}
\newcommand{\hu}{\hat{u}}
\newcommand{\hv}{\hat{v}}
\newcommand{\hw}{\hat{w}}
\newcommand{\cijk}{c_{ij}^k}
\newcommand{\cljk}{c_{lj}^k}
\renewcommand{\omega}{J}
\renewcommand{\kappa}{U}
\newcommand{\cp}{L}

\title{Quantum dynamics of Bose-Hubbard Hamiltonians beyond Hartree-Fock-Bogoliubov: The Bogoliubov backreaction approximation}

\author{I. Tikhonenkov}
\affiliation{Department of Chemistry, Ben-Gurion University of
the Negev, P.O.B. 653, Beer-Sheva 84105, Israel}
\author{J. R. Anglin}
\affiliation{Fachbereich Physik, Technische Universit\"at
Kaiserslautern, D67663, Kaiserslautern, Germany}
\author{A. Vardi}
\affiliation{Department of Chemistry, Ben-Gurion University of
the Negev, P.O.B. 653, Beer-Sheva 84105, Israel}

\begin{abstract}
We formulate a method for studying the quantum field dynamics of ultracold Bose gases
confined within optical lattice potentials, within the lowest
Bloch-band Bose-Hubbard model. Our formalism extends the two-sites results of 
Phys. Rev. Lett. {\bf86}, 000568 (2001) to the general case of $M$ lattice 
sites. The methodology is based on mapping the Bose-Hubbard Hamiltonian 
to an $SU(M)$ pseudospin problem and truncating the resulting hierarchy of 
dynamical equations for correlation functions, up to pair-correlations between
$SU(M)$ generators. Agreement with few-site exact many-particle calculations 
is consistently better than the corresponding Hartree-Fock-Bogoliubov 
approximation. Moreover, our approximation compares favorably with a more
elaborate two-particle irreducible effective action formalism, at a fraction
of the analytic and numerical effort.
\end{abstract}  
\pacs{3.75.Kk}
\maketitle

\section{\label{Sec:intro}Introduction}

Strong correlation effects, which imply enhanced quantum fluctuations around mean field order parameters, are playing an increasingly important role 
in recent experiments on dilute quantum gases.  One strategy for boosting the importance of correlations and fluctuations involves the control of coupling parameters. Interatomic interactions can be effectively tuned by means of magnetic Feshbach resonances 
\cite{Stwalley76,Tiesinga92,Tiesinga93,Timmermans99,Mies00,Inouye98}, 
allowing for a controlled transition into the
non-unitary regime $n^{1/3}a_s>1$, where the effective $s$-wave scattering length
$a_s$ is larger than the average distance between particles $n^{1/3}$ 
with $n$ being the number-density of the gas.  Quantum fluctuations also
dominate quasi-one-dimensional systems \cite{Lieb,Girardeau60,Girardeau00,
Olshanii98,Petrov00,Dunjko01,OneD,Paredes2004,Kinoshita2006} where transverse 
confinement may be used to increase the effective coupling strength 
$g_{1D}=2\hbar^2 a_s/(ml_{\perp}^2)$ without explicit control of the three-dimensional
$s$-wave scattering length. In the extreme Tonks-Girardeau strong-coupling 
regime $g_{1D} m/(\hbar^2 n)\gg 1$, spatial correlations dictate the 
impentrability of bosons, leading to ideal fermion like density distributions
\cite{Girardeau60,Girardeau00}.

An alternative to increasing effective interaction strengths, is to
decrease other (e.g. kinetic) terms in the many-body Hamiltonian. In a
Bose gas confined by an optical lattice, an effective momentum cutoff 
is introduced by controling the barrier heights, thus suppressing the
hopping frequency $J$ between adjacent sites. Given $N$ particles interacting 
with strength $U$, the strong-interaction regime is achieved for $UN/J>1$, 
as manifested in the quantum transition from a superfluid to a Mott-insulator 
phase \cite{Jaksch98,vanOosten01,Greiner02}.  

Considerable theoretical effort is currently aimed at developing efficient 
methods for the description of correlated quantum gases far from 
equilibrium. One approach relies on perturbations of the lowest-order 
mean-field theory given by the Gross-Pitaevskii (GP) equation. The result 
is a family of mean-field pairing theories. The standard 
zero-temperature Bogoliubov prescription \cite{Bogoliubov47}
gives the natural small-oscillation 
modes by linearization about the GP ground-state. However, this linear response 
theory does not account for the backreaction of excitations on the 
condensate order-parameter and is thus limited to small perturbations and
short timescales. Backreaction is accounted for within the 
Hartree-Fock-Bogoliubov (HFB) theory, which prescribes a set of coupled
equations for the condensate order-parameter and pair correlation functions
\cite{Griffin96,Proukakis1,Holland01,Rey04}. 
Since both normal and anomalous correlations are included, this approach comes 
at the cost of ultraviolet divergences of anomalous quantities. While this problem is 
relatively easy to deal with by renormalization of the coupling parameters, 
a more serious issue, also related to the inclusion of anomalous correlations, 
is the HFB spectral gap \cite{Griffin96}. This unphysical gap in the 
excitation spectrum 
results in from the breaking of $U(1)$ gauge symmetry and the consequent 
elimination of the Goldstone modes corresponding to gauge transformations 
of the broken symmetry solution. An intermediate form between Bogoliubov and
HFB is the HFB-Popov (HFB-P) approximation \cite{Griffin96,Popov} where $U(1)$ 
symmetry is restored by
elimination of {\it noncondensate} anomalous terms only. While the resulting 
theory is gapless, it does not conserve the total number of particles and is 
thus inadequate for describing dynamical condensate depletion. Finally, 
if all anomalous quantities are neglected, one obtains the bosonic Hartree 
Fock (HF) theory \cite{Griffin96} which is both gapless and conserving, 
but does not allow for any dynamical depletion, since the populations of condensed 
and non-condensed particles are conserved separately. It is thus highly desirable to develop 
a theoretical description that (a) is $U(1)$ invariant and hence gapless,  
(b) conserves the total number of particles, yet (c) allows for dynamical 
depletion of the condensate.

Recently, a perturbative approximation scheme based on a two-particle 
irreducible (2PI) effective action expansion, has been used to study the 
nonequilibrium dynamics of condenstaes in optical lattices \cite{Rey04}
within the lowest Bloch-band Bose-Hubbard model. Within the 2PI effective 
action expansion, the Bogoliubov and HFB theories emerge as one-loop and
a single two-loop correction respectively, to the classical GP action.
Higher-order approximations, obtained by including two-vertex terms in the
diagramatic expansion of the effective action (denoted as in Ref. \cite{Rey04} 
by '2nd')  and by a $1/{\cal N}$ expansion up to second-order in the
coupling strength (denoted henceforth by '$1/{\cal N}$') with ${\cal N}$ 
being the number of auxillary classical fields used to approximate the 
quantum-field, have been compared with HFB and exact few-sites numerical 
calculations. The results demonstrate some improvement of the higher-order approximations over HFB in predicting the exact many-body dynamics. However, at sufficiently long times 
all approximations fail due to interaction effects. 
A nonperturbative $1/{\cal N}$ 2PI effective action expansion approach have also been 
developed and applied to the equilibration of a homogeneous Bose gas in 1D \cite{Gasenzer05}.

In this work we develop a mean-field theory for the description of
quantum dynamics in the Bose-Hubbard model. The technique, referred to
here as Bogoliubov Back Reaction (BBR), is a many-site extension of 
previous work on a two-site model \cite{Vardi01,Anglin01}, based on 
the perturbation of equations of motion for the reduced single-particle 
density operator, instead of the usual field operator approach. The resulting
equations involve the two-point reduced single-particle density matrix (SPDM)
and the four-point correlation functions.  They contain only normal 
({\it i.e.}  number conserving) quantities, and are thus $U(1)$ symmetric. 
The approximation conserves the total 
number of particles, yet it allows for population transfer from the
condensate to the excitations, thus accounting for condensate depletion
during the evolution. We compare BBR calculations with full many-body 
numerical results for up to a hundred particles and five lattice sites,
as well as with HFB and 2PI effective action results. The BBR results
give better, longer-time predictions than current rival approximations, at
a small fraction of the theoretical effort.  

In section II we present the Bose-Hubbard model and the standard HFB
approach. In section III we transform the Bose-Hubbard Hamiltonian 
with $M$ lattice sites into an $SU(M)$ pseudospin problem, derive 
dynamical equations for 
the $SU(M)$ generators spanning the single-particle density operator,
and truncate the resulting hierarchy of dynamical equations for correlation 
functions to obtain the BBR equations of motion. Section IV contains
numerical few-sites results and comparison with HFB as well as 
2PI effective action approximation methods. Discussion, conclusions and 
prospects for future research are presented in section V.

\begin{figure}
\centering
\includegraphics[width=0.5\textwidth]{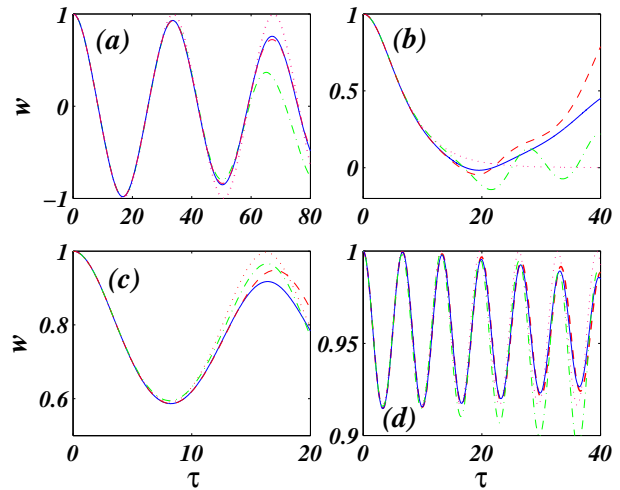}
\caption{(color online) Population imbalance $w$ in a two-site system 
as a function of rescaled
time $\tau$ for $L=2$ (a), $L=4$ (b), $L=5$ (c), and $L=10$ (d). The
total number of particles is set to $N=100$. Solid blue
lines, corresponding to exact many-body numerical results, are compared to
the GP (dotted lines), HFB (dash-dotted lines), and BBR (dashed lines) 
approximations.} 
\label{fig1}
\end{figure}   
 
\section{Conventional Mean-Field Theories: Gross-Pitaevskii and Hartree-Fock-Bogoliubov}
We begin with the standard Bose-Hubbard model Hamiltonian for 
an ultracold gas in a one-dimensional periodic optical lattice
\begin{equation}
\label{Ham}
\hat{H}=\omega\sum_i \left(\ha_{i+1}^\dag\ha_i+\ha_i^\dag\ha_{i+1}\right)
+\frac{\kappa}{2}\sum_i \ha_i^\dag\ha_i^\dag\ha_i\ha_i
\end{equation}
where $\ha_i$ and $\ha^\dag_i$ are annihilation and creation operators 
respectively, for a particle in site $i$. We consider only on-site interactions 
with strength $\kappa$ and nearest-neighbor tunneling with hopping rate
$\omega$. These approximations are justified because adjacent site interactions 
and next-to-nearest-neigbhbor tunneling amplitudes are characteristically
at least two orders of magnitude smaller than on-site interactions and 
nearest-neighbor hopping \cite{Jaksch98}. The Bose-Hubbard model is 
viable as long as there are no transitions into excited Bloch bands.   

Using the Hamiltonian (\ref{Ham}) we write the Heisenberg equations of motion
for the field-operators $\ha_j$, 
\begin{equation}
i\ddt\ha_j=\omega\left(\ha_{j-1}+\ha_{j+1}\right)
+\kappa\ha_j^\dag\ha_j\ha_j~.
\label{aeom}
\end{equation}
The lowest order mean field theory for the Bose-Hubbard model is
obtained by replacing the field operators $\ha_j$, and $\ha_j^\dag$
the by $c$-numbers $a_j$ and $a_j^*$. This approximation is tantamount
to assuming coherent many-body states with a well-defined phase 
between sites. Rescaling $a\rightarrow\sqrt{N}a$ and $\tau=Jt$ 
we arrive at the discrete GP equation,
\begin{equation}
i\ddtau a_j=\left(a_{j-1}+a_{j+1}\right)
+\cp |a_j|^2 a_j~,
\label{gp}
\end{equation}
where $\cp=\kappa N /\omega$ is the characteristic coupling parameter.
Within the GP mean field theory (\ref{gp}) fluctuations are completely 
neglected and the system is always assumed to be described by a single, 
coherent order parameter. Therefore an accurate description of the superfluid 
to Mott insulator quantum phase transition is not possible. Nevetheless, 
qualitative differences exist between mean field dynamics in the weak-coupling
regime $L<2$, where the system exhibits full-amplitude Rabi-like oscillations,
and the strong coupling case $L>2$, where self-trapped motion is observed 
\cite{Smerzi97,Raghavan99,Vardi01,Anglin01,Rey04}

To go beyond the GP approximation, a higher-order mean field theory may
be formulated by adding to Eq. (\ref{aeom}) additional equations of 
motion for the normal density operators $\ha_j^\dag\ha_k$, and the anomalous 
density operators $\ha_j\ha_k$,
\begin{eqnarray}
\label{ddto}
i\ddt\ha_j\ha_k&=&\omega\left(\ha_k\ha_{j-1}+\ha_k\ha_{j+1}
+\ha_j\ha_{k-1}+\ha_j\ha_{k+1}\right)\nonumber\\
\label{ddtnor}
~&~&+\kappa\left(\ha_j^\dag\ha_k\ha_j\ha_j
+\ha_k^\dag\ha_j\ha_k\ha_k\right)\nonumber\\
~&~&+\frac{\kappa}{2}\left(\ha_j\ha_j+\ha_k\ha_k\right)\delta_{jk}~,\\
\label{ddtan}
i\ddt\ha_j^\dag\ha_k&=&\omega\left(\ha_j^\dag\ha_{k-1}+\ha_j^\dag\ha_{k+1}
-\ha_{j-1}^\dag\ha_k-\ha_{j+1}^\dag\ha_k\right)\nonumber\\
~&~&+\kappa\left(\ha_j^\dag\ha_k^\dag\ha_k\ha_k
-\ha_j^\dag\ha_j^\dag\ha_j\ha_k\right)~.
\end{eqnarray}
Taking the expectation values of Eq. (\ref{aeom}) and 
Eqs. (\ref{ddto})-(\ref{ddtan}), and using the HFB Gaussian ansatz, 
we truncate third- and fourth-order moments as:
\begin{eqnarray} 
\langle \hA\hB\hC \rangle&\approx& \langle\hA\rangle\langle\hB\hC\rangle+
\langle\hB\rangle\langle\hA\hC\rangle\nonumber\\
~&~&+\langle\hC\rangle\langle\hA\hB\rangle
-2\langle\hA\rangle\langle\hB\rangle\langle\hC\rangle~,\\
\langle \hA\hB\hC\hD \rangle&\approx& 
\langle\hA\hB\rangle\langle\hC\hD\rangle
+\langle\hA\hC\rangle\langle\hB\hD\rangle\nonumber\\
~&~&+\langle\hA\hD\rangle\langle\hB\hC\rangle
-2\langle\hA\rangle\langle\hB\rangle\langle\hC\rangle\langle\hD\rangle~,
\end{eqnarray}
to obtain the HFB equations:
\begin{eqnarray}
\label{mfhfb}
i\ddtau\ha_j&=&\left(a_{j-1}+a_{j+1}\right)\nonumber\\
~&~&+\cp a_j^*a_ja_j+\cp\left(2 a_j\dnor_{jj}+ a_j^*\dan_{jj}\right)~,\\
\label{anhfb}
i\ddtau\dan_{jk}&=&\left(\dan_{j-1,k}+\dan_{j+1,k}
+\dan_{j,k-1}+\dan_{j,k+1}\right)\nonumber\\
~&~&+2\cp\left(|a_j|^2+|a_k|^2+\dnor_{jj}
+\dnor_{kk}\right)\dan_{jk}\nonumber\\
~&~&+\frac{\cp}{2}\left(a_k^2+\dan_{kk}\right)\left(2{\dnor_{jk}}^*
+\delta_{jk}\right)\nonumber\\
~&~&+\frac{\cp}{2}\left(a_j^2+\dan_{jj}\right)\left(2\dnor_{jk}
+\delta_{jk}\right)\\
\label{norhfb}
i\ddtau\dnor_{jk}&=&\left(\dnor_{j,k-1}+\dnor_{j,k+1}
-\dnor_{j-1,k}-\dnor_{j+1,k}\right)\\
~&~&+2\cp\left[\left(|a_k|^2+\dnor_{kk}\right)
-\left(|a_j|^2+\dnor_{jj}\right)\right]\dnor_{jk}\nonumber\\
~&~&+\cp\left[\left(a_k^2+\dan_{kk}\right){\dan_{jk}}^*-
\left(a_j^2+\dan_{jj}\right)^*\dan_{jk}\right]~,\nonumber
\end{eqnarray}
for the mean field $a_j\equiv\langle\ha_j\rangle/\sqrt{N}$ and the two-point 
correlation functions
$\dnor_{jk}=[\langle\ha_j^\dag\ha_k\rangle-a_j^*a_k]/N$, 
$\dan_{jk}\equiv\left[\langle\ha_j\ha_k\rangle-a_j a_k\right]/N$, 
constituting the reduced single particle density matrix. 

We note that
the discrete HFB equations (\ref{mfhfb})-(\ref{norhfb}) are not UV divergent 
due to the natural momentum cutoff imposed by the lattice. However, 
due to the existence of a noncondensate anomalous density, $U(1)$ symmetry
is broken, in contrast to the gauge-invariant original 
field equations (\ref{aeom}). $U(1)$ symmetry may be restored for example,
by ommitting all anomalous quantities, to obtain the Hartree-Fock equations
\begin{eqnarray}
\label{sfhf}
i\ddtau a_j&=&\left(a_{j-1}+a_{j+1}\right)+\cp\left(|a_j|^2+2\dnor_{jj}\right)a_j~,\\
\label{norhf}
i\ddtau\dnor_{jk}&=&\left(\dnor_{j,k-1}+\dnor_{j,k+1}
-\dnor_{j-1,k}-\dnor_{j+1,k}\right)\\
~&~&+2\cp\left[\left(|a_k|^2+\dnor_{kk}\right)
-\left(|a_j|^2+\dnor_{jj}\right)\right]\dnor_{jk}.\nonumber
\end{eqnarray}  
Equations (\ref{sfhf}) and (\ref{norhf}), conserve separately the 
condensate population $\sum_j |a_j|^2$ and the noncondensed fraction 
$\sum_j \dnor_{jj}$. Thus, the HF approximation can not be used to account 
for condensate depletion during the evolution. If only the noncondensate
anomalous terms are neglected, one obtains the HFB-Popov \cite{Popov} 
approximation, which allows for growth of fluctuations, but conserves
the condensate population, so that the total number is not a constant
of motion. In the following section we construct a $U(1)$ invariant 
mean-field theory which conserves the total number of particles, 
yet includes dynamical depletion. 

\begin{figure}
\centering
\includegraphics[width=0.5\textwidth]{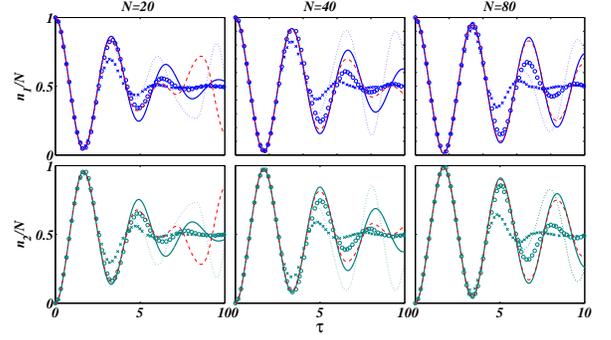}
\caption{(color online) Evolution of atomic site populations in a two-site
system, starting with all population in one site, for $N=20,40,80$ and fixed $L=2$. 
Exact numerical results (solid)
are compared with the HFB (dotted) and BBR (red dashed lines) approximations,
as well as to the two approximations based on the 2PI effective action 
formalism: 2nd order (x's) and $1/{\cal N}$ (circles), taken from Fig. 5 in
Ref. \cite{Rey04}.}
\label{fig2}
\end{figure}

\begin{figure}
\centering
\includegraphics[width=0.5\textwidth]{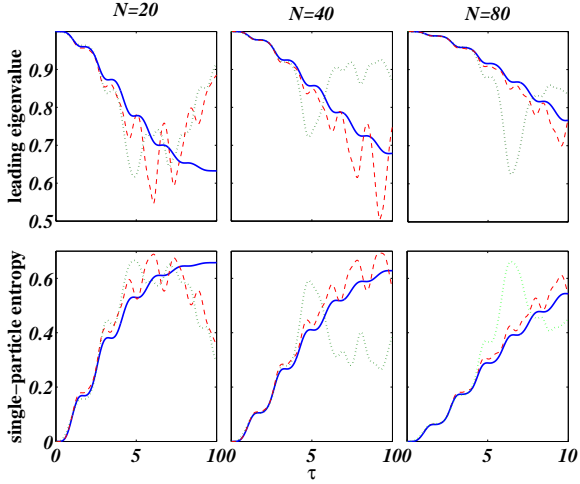}
\caption{(color online) Evolution of the leading eigenvalue (above) and 
single-particle entropy (below) for a two-site system with $N=20,40,80$
and $L=2$. Exact many-body numerics (solid blue line) is compared with the HFB 
approximation (green dotted line) and the BBR approximation (red dashed
line).}
\label{fig3}
\end{figure}    

\section{The Bogoliubov Backreaction Equations}
Instead of the conventional mean-field approaches, based on
the site field operators $\ha_j$, we construct a mean field formalism
using the reduced single-particle density operator $\ha_j^\dag\ha_k$,
treating it as the fundamental quantity. We have previously applied this 
approach to the case of a two-site model \cite{Vardi01,Anglin01}. Here we
extend it to the general $M$ site case. It is convenient to rewrite 
the Hamiltonian (\ref{Ham}) in terms of the $M^2-1$ traceless operators  
which generate $SU(M)$:
\begin{eqnarray}
\hu_{j,k}&=&\ha_j^\dag\ha_k+\ha_k^\dag\ha_j,~1\le k<j\le M\nonumber\\
\hv_{j,k}&=&-i\left(\ha_j^\dag\ha_k-\ha_k^\dag\ha_j\right),~1\le k<j\le M\\
\hw_l&=&-\sqrt{\frac{2}{l(l+1)}}\left(\sum_{j=1}^l\hn_j-l\hn_{l+1}\right),~1\le l\le M-1~.\nonumber
\end{eqnarray} 
Since it is easily verified that 
\begin{equation}
\frac{1}{2}\sum_{j=1}^{M-1} \hw_j^2+\frac{1}{M}\hn^2=\sum_{j=1}^{M}\hn_j^2
\end{equation}
where $\hn=\sum_{j=1}^{M}\hn_j$ is the total particle number, equation (\ref{Ham}) can be
rewritten, eliminating $c$-number terms, as:
\begin{equation}
\hat{H}=\omega\sum_{j=1}^{M-1}\hu_{j+1,j}+\frac{\kappa}{4}\sum_{j=1}^{M-1} \hw_j^2~.
\label{hsun}
\end{equation}

Using the $SU(M)$ generators we construct a pseudospin vector operator,
\begin{eqnarray}
{\hat{\bf S}}=&\left(\hu_{21},\hu_{32},\dots,\hu_{31},\hu_{42},\dots,\hv_{21},\hv_{32},\right.\nonumber\\
~&\left .\dots,\hv_{31},\hv_{42},\dots,\hw_1,\hw_2,\dots,\hw_{M-1}\right),
\end{eqnarray} 
so that the Hamiltonian (\ref{hsun}) takes the form:
\begin{equation}
\label{Hamsun}
\hat{H}=\omega\sum_{j=1}^{M-1}\hS_j+\frac{\kappa}{4}\sum_{j=M^2-M}^{M^2-1} \hS_j^2.
\end{equation} 
The Heisenberg equations of motion for the operators $\hS_i$ and their products $\hS_i\hS_l$ then read:
\begin{eqnarray}
\label{oseom}
i\ddt\hS_i&=&\omega\sum_{j=1}^{M-1}\sum_{k=1}^{M^2-1}\cijk\hS_k\\
~&~&+\frac{\kappa}{4}\sum_{j=M^2-M+1}^{M^2-1}\sum_{k=1}^{M^2-1}\cijk(\hS_k\hS_j+\hS_j\hS_k),\nonumber\\
\label{osseom}
i\ddt\hS_i\hS_l&=&\omega\sum_{j=1}^{M-1}\sum_{k=1}^{M^2-1}(\cijk\hS_k\hS_l+\cljk\hS_i\hS_k)\\
~&~&+\frac{\kappa}{4}\sum_{j=M^2-M+1}^{M^2-1}\sum_{k=1}^{M^2-1}\cijk(\hS_k\hS_j+\hS_j\hS_k)\hS_l\nonumber\\
~&~&+\frac{\kappa}{4}\sum_{j=M^2-M+1}^{M^2-1}\sum_{k=1}^{M^2-1}\cljk\hS_i(\hS_k\hS_j+\hS_j\hS_k),\nonumber
\end{eqnarray}
where the coefficients $\cijk$ are the structure constants of the $SU(M)$ group. We note that for $M=2$ the Hamiltonian (\ref{hsun}) and the dynamical equations (\ref{oseom})-(\ref{osseom}) reduce to the familiar Bloch forms used in Refs. \cite{Vardi01,Anglin01}. The $M$-site system is a direct extension of the two-mode case, in that hopping terms induce linear Rabi-like oscillations in the ${\bf vw}$ subspace, whereas on-site interactions lead to nonlinear phase precession in the ${\bf uv}$ subspace.

\begin{figure}
\centering
\includegraphics[width=0.5\textwidth]{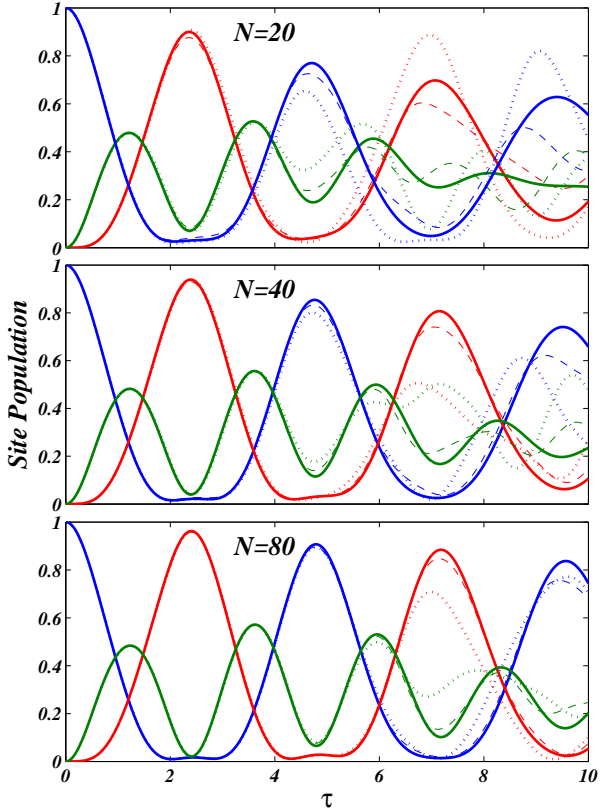}
\caption{(color online) Site-populations in a three-site system 
as a function of rescaled time $\tau$
for $N=20,40,80$ and fixed $L=2$. Blue, green, and red lines correspond to
1st, 2nd, and 3rd site populations, respectively. Solid lines depict the
full many-body dynamics, whereas dotted and dashed lines correspond to the
HFB approximation and the BBR approximation, respectively. 
}
\label{fig4}
\end{figure}

\begin{figure}
\centering
\includegraphics[width=0.5\textwidth]{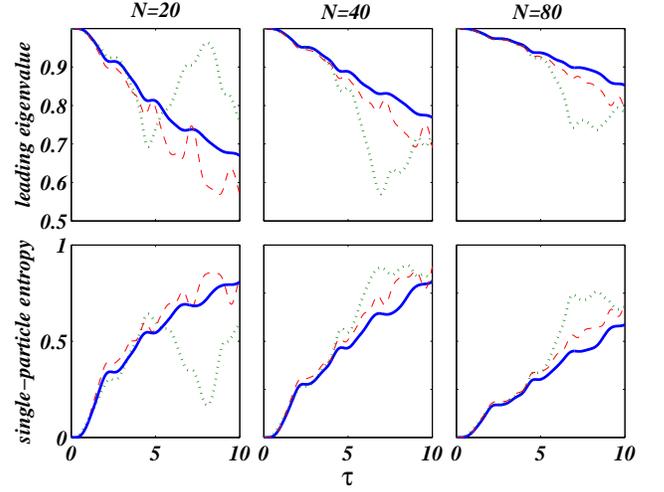}
\caption{(color online) Leading eigenvalue of the reduced single-particle density
matrix $\rho$ and single-particle entropy $-{\rm Tr}(\rho\ln\rho)$, as a function of
rescaled time $\tau$ in a three-site system with $N=20,40,80$ and $L=2$. Exact results
(solid blue lines) are compared to HFB calculations (dotted green lines) and BBR 
calculations (dshed red lines).}
\label{fig5}
\end{figure}

The reduced single-particle density matrix is obtained from the expectation value
of ${\hat{\bf S}}$, according to:
\begin{equation}
\rho=\frac{N}{2}{\cal I}+\frac{1}{2}\sum_{j=1}^{M^2-1} \langle\hS_j\rangle \sigma_j~,
\end{equation}
where ${\cal I}$ is a unit matrix of order $M$ and $\sigma_j$ are the $M\times M$
irreducible representations of the $SU(M)$ generators (e.g. Pauli matrices
for $M=2$, Schwinger matrices for $M=3$ etc.). We will therefore focus on
the dynamics of the 'hyper-Bloch-vector' 
${\bf S}\equiv\langle{\hat{\bf S}}\rangle/2N$. 
The lowest-order mean-field approximation replaces the vector of operators 
${\hat{\bf S}}$ by the vector of their expectation values 
${\bf S}$, thus truncating 
$\langle \hS_i\hS_j\rangle\approx\langle\hS_i\rangle\langle\hS_j\rangle$.
This results in the nonlinear pseudospin-precession form of the GP equations,
\begin{equation}
\ddtau{\bf S}={\bf B}({\bf S})\otimes{\bf S}
\label{sprec}
\end{equation}
where
\begin{equation}
{\bf B}({\bf S})=\left(B_1,B_2,\dots,B_{M^2-1}\right)~,
\end{equation}
with
\begin{equation}
B_j=\left\{\begin{array}{ll}
1&j=1,\dots,M(M-1)/2\\
0&j=M(M-1)/2+1,\dots, M(M-1)\\
\cp S_j&j=M(M-1)+1,\dots,M^2-1\end{array}\right.
\nonumber
\end{equation}

It is readily verified that Eq. (\ref{sprec}) is exactly equivalent to the
discrete GP equation (\ref{gp}). In addition to the conservation of the
total number ${\rm Tr}(\rho)$ there exist, within GP theory, $M-1$ independent
constants of the motion ${\rm Tr}(\rho^m)$ with $m=2,\dots,M-1$. For example, for 
$M=2$ the GP mean-field theory also conserves the single-particle purity 
${\rm Tr}(\rho^2)$, which is just the length of the three-dimensional 
Bloch vector. Deviations from this classical field theory, due to interparticle
entanglement and loss of single particle coherence, will show up as
a reduction in these classically conserved quantities.

The BBR approximation is obtained by going one level deeper in the hierarchy
of dynamical equations for expectation values. Taking the expectation values 
of Eq. (\ref{oseom}) and Eq. (\ref{osseom}) and truncating 
\begin{eqnarray} 
\langle \hS_i\hS_j\hS_k \rangle&\approx& \langle\hS_i\rangle\langle\hS_j\hS_k\rangle+
\langle\hS_j\rangle\langle\hS_i\hS_k\rangle\nonumber\\
~&~&+\langle\hS_k\rangle\langle\hS_i\hS_j\rangle
-2\langle\hS_i\rangle\langle\hS_j\rangle\langle\hS_k\rangle~,
\end{eqnarray} 
we obtain the BBR equations of motion:
\begin{eqnarray}
i\ddtau S_i&=&\sum_{j=1}^{M-1}\sum_{k=1}^{M^2-1}\cijk S_k\nonumber\\
~&~&+\cp\sum_{j=M^2-M+1}^{M^2-1}\sum_{k=1}^{M^2-1}\cijk (S_j S_k+\Delta_{jk})\\
i\ddtau\Delta_{il}&=&\sum_{j=1}^{M-1}\sum_{k=1}^{M^2-1}\left(\cijk\Delta_{lk}+\cljk\Delta_{ik}\right)\\
~&~&+\cp\sum_{j=M^2-M+1}^{M^2-1}\sum_{k=1}^{M^2-1}\cijk\left(\Delta_{lj}S_k+\Delta_{lk}S_j\right)\nonumber\\
~&~&+\cp\sum_{j=M^2-M+1}^{M^2-1}\sum_{k=1}^{M^2-1}\cljk\left(\Delta_{ij}S_k+\Delta_{ik}S_j\right),\nonumber
\end{eqnarray}
where $S_j=\frac{\langle \hS_j\rangle}{2N}$ and $\Delta_{jk}=\frac{\langle \hS_j\hS_k+\hS_k\hS_j\rangle-2S_jS_k}{4N^2}$. In the following section we compare the accuracy of the BBR approximation with respect to GP, HFB, and 2PI effective action.

\begin{figure}
\centering
\includegraphics[width=0.5\textwidth]{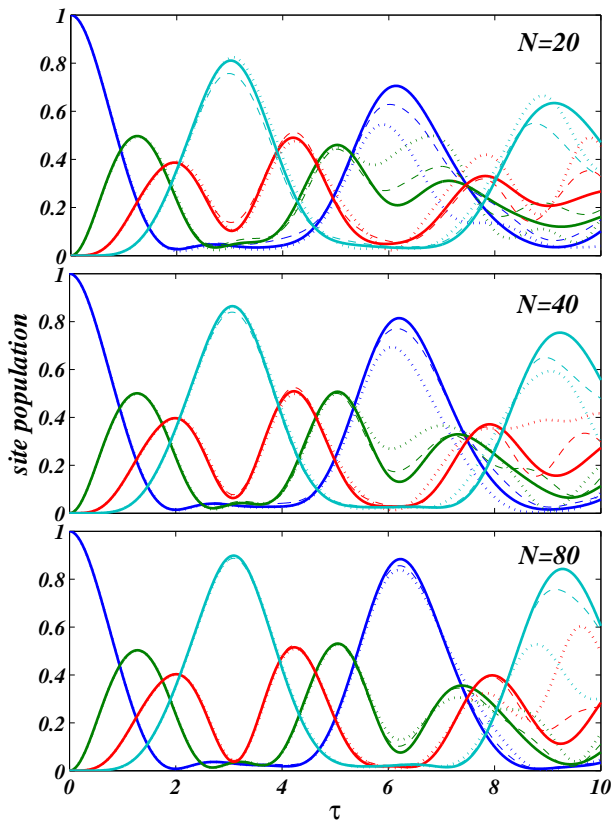}
\caption{(color online) Site-populations in a four-site system 
as a function of rescaled time $\tau$
for $N=20,40,80$ and fixed $L=2$. Blue, green, red, and cyan lines correspond to
1st, 2nd, 3rd, and 4th site populations, respectively. Solid lines depict the
full many-body dynamics, whereas dotted and dashed lines correspond to the
HFB approximation and the BBR approximation, respectively. 
}
\label{fig6}
\end{figure}

\section{Numerical results}

In order to test the accuracy of the BBR approximation compared to other
methods, we carried out exact numerical calculations for limited numbers
of particles and sites (up to $N=100$ particles and $M=5$ lattice sites).
The Hamiltonian (\ref{Ham}) was represented in terms of site-number states
and the $N$-body Schr{\"o}dinger equation was solved numerically, as in 
Refs. \cite{Rey04,Vardi01,Anglin01}. These many-body results were then
compared with BBR mean-field calculations, as well as with GP, HFB and 
variants of the 2PI effective action method. 

In Fig \ref{fig1} the evolution of fractional population difference for 
a hundred particles in two-sites, is plotted for various values of the
coupling parameter $\cp$. Within the GP mean-field theory, full-amplitude
Rabi-like oscillations are predicted in the linear regime with $L<2$ 
(Fig \ref{fig1}(a)). As the transition is made to the strong-coupling regime, 
the oscillation becomes increasingly more nonlinear, until when $L\ge4$ 
macroscopic self-trapping is attained  (Figs. \ref{fig1}(b)-\ref{fig1}(d)).
The value of $L=4$ is particularly interesting because for this coupling
a trajectory starting from a single-populated site becomes dynamically 
unstable when site-populations equilibrate. In previous work 
we have shown that this dynamical instability serves as a quantum-noise 
amplifier \cite{Vardi01,Anglin01}, so that the growth of the deviation 
of a quantum trajectory from the corresponding GP prediction is initially 
exponential, leading to a $log(1/N)$ slow convergence of the many-body 
quantum-field results to the classical GP prediction. Thus, while the
naive expectation would be that quantum fluctuations would simply grow
with the coupling parameter $L$, their role is in fact maximized for $L=4$,
as evident in  Fig. \ref{fig1}(b). It is clear from Fig. \ref{fig1} that
the BBR approximation gives a better description of the ensuing quantum dynamics, for longer timescales, than HFB does.

Convergence of various approximations with increasing number of particles
is demonstrated in Fig. \ref{fig2}, where the two-sites population 
dynamics is plotted for increasing particle numbers, keeping a fixed
coupling value of $L=2$. In addition to the exact, BBR, and HFB results,
we also plot two calculations based on the 2PI effective action approach,
taken from Fig. 5 of Ref. \cite{Rey04} (our exact and HFB results 
exactly coincide with the corresponding lines in that figure). Here
too, the BBR approximation (red dashed lines) gives a more accurate 
description of the dynamics than any of the other methods, attaining a
nearly perfect convergence in the given time-frame for $N=80$ particles.
In comparison, standard HFB fails to depict the damping of coherent 
oscillations, whereas the 2PI effective action methods tend to overdamp.
We note, that in terms of formalistic complexity alone, the BBR 
approximation is far simpler than the noninstantaneous integrodifferential
equations used in the 2PI effective action methods \cite{Rey04}. In fact,
it is even simpler than HFB, in that only normal quantitities are involved,
giving a total of nine equations for two sites, as opposed to fifteen in HFB.

Dynamical condensate depletion is also well-depicted by the BBR approximation.
In Fig. \ref{fig3} we plot the evolution of the leading eigenvalue of the 
reduced single-particle density matrix $\rho$ and the single-particle  
von-Neumann entropy $-{\rm Tr}(\rho\ln\rho)$, corresponding to the population 
dynamics of Fig. \ref{fig2}. While HFB calculations seem to give an abrupt
deviation of the predicted condensate fraction from its exact value, the 
BBR results converge well, giving a reasonably accurate description of
BEC depletion.   
 
The same qualitative behavior carries over to systems with more than two sites.
In Fig.~\ref{fig4}  and Fig.~\ref{fig5}, population dynamics and condensate 
depletion are shown for a three-sites system with $N=20,40,80$ particles. 
Similarly to the two-sites case, the BBR approximation constitutes a significant
improvement over the HFB approach, giving a better description of populations
as well as coherences. The same is also true for the four-sites case
shown in Fig. \ref{fig6}.
 
\begin{figure}
\centering
\includegraphics[width=0.5\textwidth]{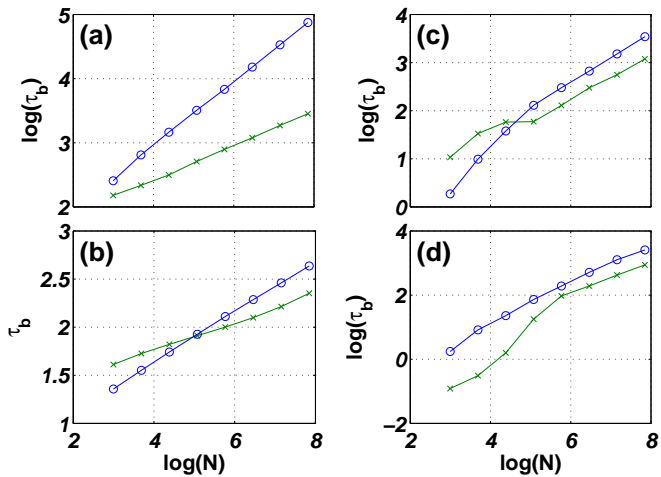}
\caption{(color online) Characteristic times at which the Cartesian distance
between the exact Bloch vector and its HFB (green, x's) and BBR (blue, circles)
approximants, reaches a predtermined threshold, as a function of $N$ for $\cp=1$ 
(a), $\cp=4$ (b), $\cp=6$ (c), and $\cp=10$ (d). The break-threshold is set to 
$0.2$ in (a)-(c), and to 0.05 in (d).}
\label{fig7}
\end{figure}

The faster convergence of BBR as compared with the HFB appoximation is illustrated in
Fig. \ref{fig7}, where characteristic breaktimes of the two approximations in a
two-sites calculation, are plotted as a function of the total number of particles $N$. 
As anticipated, breaktimes grow as $\sqrt{N}$ when the classical dynamics is regular 
(\ref{fig7}a,\ref{fig7}c,\ref{fig7}d) and as $\log{N}$ when the classical 
trajectory hits the dynamical instability (\ref{fig7}b). The BBR calculations
give consistently longer breaktimes, with a more regular convergence pattern.

\section{Discussion}

The rich regime of strongly correlated many body physics, which ultracold atom experiments are now beginning to probe, will surely not be fully conquered by any simple hierarchy truncation scheme such as BBR.  Nor does BBR offer anything like an exact solution even to the problems to which we have applied it in this paper; its improvements over its rivals are incremental rather than revolutionary.  On the other hand it should be born in mind that incremental improvements in theory are more significant in the context of ultracold gases than in traditional condensed matter, because in the new atomic systems samples are precisely characterized, controlled, and measured, and relevant microphysics is clearly known.  It is perfectly plausible in these systems that we may come to learn important qualitative principles from experimental discrepancies on the few percent level.

The merits of BBR that we would like to emphasize, along with its very reasonable level of accuracy, are its simplicity and its direct relation to experimental reality.  It involves only quantities which are directly observed in single- and two-particle number-conserving measurements, and it respects the fact that in current quantum gas laboratories atoms are neither created nor destroyed.  And it is conceptually and computationally straightforward.

In one sense it is of course conceptually all too straightforward: like all hierarchy truncation schemes since Boltzmann's, it is an uncontrolled approximation, whose accuracy is therefore arguably as much a puzzle as it is a solution.  Insofar as truncating a hierarchy at a deeper level is grounds for expecting higher accuracy, however, the advantage of BBR is clear: it is a truncation at fourth order in field operators, compared to only second order for HFB.  Deeper level truncation often involves proliferation of terms, to the point of sharply diminishing returns in accuracy versus effort; but BBR avoids this, and manages to use fewer equations than HFB, because it eliminates all anomalous terms.

And this leads us to conclude by indicating some of the potential future applications of the results of this paper.

Why do hierarchy truncations often work as well as they do?  What determines the best way of truncating a hierarchy?  These are questions that have been raised ever since Boltzmann's {\it stosszahlansatz} produced the arrow of time, but they have yet to be fully answered.  With current experimental capabilities for precise and controlled measurements on ultracold gases, introducing a physically motivated alternative truncation scheme, as this paper has done, may contribute to new progress on these questions.

Finally, another conceptual merit of BBR is that because it is based on the single particle density matrix, rather than just the macroscopic wave function, it makes such a conceptually important quantity as the single particle entropy -- the entropy of Boltzmann -- a basic ingredient in the theory, rather than a perturbative afterthought.  Rethinking entropy, heretofore mainly in the context of quantum information and computation theory, is another major thrust of current physics; the alternative viewpoint offered by BBR may potentially be of some value in a broader conception of this project.

\begin{acknowledgements}
This work was supported in part by grants from the Minerva foundation
for a junior research group and the Israel Science Foundation for a
Center of Excellence (grant No.~8006/03).
\end{acknowledgements}

\end{document}